\documentclass[letter,twocolumn]{jpsj2}

\usepackage{graphicx}

\def\ve#1{\mbox{\boldmath $#1$}}

\def\Tc{T_\mathrm{c}}
\def\Re{\mathrm{Re}}

\title{%
Analysis of Superconductivity in 
d-p Model on Basis of Perturbation Theory\\
}

\author{%
Sotaro \textsc{Sasaki},
Hiroaki \textsc{Ikeda}
and Kosaku \textsc{Yamada}
}

\inst{
Department of Physics, Kyoto University, Sakyo-ku, Kyoto 606-8502 \\
}

\recdate{\today}

\abst{%
We investigate the mass enhancement factor and the superconducting transition temperature in the d-p model 
for the high-$\Tc$ cuprates. 
We solve the \'Eliashberg equation using the third-order perturbation theory 
with respect to the on-site Coulomb repulsion $U$. 
We find that when the energy difference between the d-level and p-level is large, 
the mass enhancement factor becomes large and $\Tc$ tends to be suppressed 
owing to the difference of the density of state for d-electron at the Fermi 
level. 
From another viewpoint, 
when the energy difference is large, the d-hole number approaches 
unity and the electron correlation becomes strong 
and enhances the effective mass. This behavior for the electron number 
is the same as that for the 
f-electron number in the heavy fermion systems. 
The mass enhancement factor plays an essential role in understanding the 
difference in $\Tc$ between the LSCO and YBCO systems. 
}

\kword{%
d-p model, high-$\Tc$ cuprate, mass enhancement factor, transition temperature, third-order perturbation theory
}

\begin{document}
\sloppy
\maketitle

Currently, superconductivity in high-$\Tc$ cuprates is being intensively investigated. 
For example, from the point of view of strong coupling, 
the $t$-$J$ model has been investigated. ~\cite{rf:Lee}
Also, Hubbard model has been investigated using the variational Monte Carlo 
method~\cite{rf:Nakanishi} and the quantum Monte Carlo method.~\cite{rf:Kuroki} On the other hand, from the point of view of weak coupling, 
the Hubbard model has been investigated using the fluctuation-exchange approximation and the perturbation theory. In the latter case, 
the nature of superconductivity in the high-$\Tc$ cuprates has almost been clarified on the basis of the nearly 
antiferromagnetic Fermi-liquid theory.~\cite{rf:Moriya,rf:Yanase} 
However, we have not yet explained the observed differences between the high-$\Tc$ systems, 
YBa$_2$Cu$_3$O$_{7-\delta}$(YBCO) and La$_{2-x}$Sr$_x$CuO$_4$(LSCO) and so on, 
in particular, the principal reason why the transition temperature $\Tc$ observed 
in LSCO is relatively low. 
The nuclear quadrupole resonance (NQR) experiment by Zheng {\it et al}. shows that when the ratio of the 
d-hole number ($n_d$) to the p-hole number ($n_p$), $n_d/2n_p$, is large, 
$\Tc$ 
is suppressed. ~\cite{rf:Zheng}
Actually, LSCO has a large ratio, as compared with YBCO. 
Also, in the specific heat experiment, 
the $\gamma$-value per mole in LSCO 
is approximately as large as that in 
YBCO. ~\cite{rf:Loram,rf:Momono}
Considering that YBCO has two CuO$_2$ layers and three Cu atoms 
in the unit, we find that the 
$\gamma$-value per layer in LSCO is large. 
The effect of the strong mass enhancement can be also seen in the nuclear magnetic resonance (NMR) relaxation rate. 
From the  systematic study of $(T_1T)^{-1}$, we can see that the effective Fermi energy in LSCO is rather smaller than that in YBCO.~\cite{rf:Kitaoka}
Thus, the strong electron correlation leads to a large mass enhancement, 
and then the $\Tc$ in LSCO is reduced. 
This proposal is very important to clarify a relation to the other materials, 
such as the heavy fermion systems, and to promote further progress on the 
unified picture in the strongly correlated systems. 
In this study, using the third-order perturbation theory, 
we investigate the mass enhancement factor and $\Tc$ in the d-p model 
for high-$\Tc$ cuprates, and clarify the difference in $\Tc$ between LSCO and 
YBCO. 
Husimi discussed it in the same framework and concluded that it 
originates from the difference in the effective on-site Coulomb repulsion 
$U$.~\cite{rf:Husimi} 
We would like to stress here that the mass enhancement factor plays an 
essential role. 

\begin{figure}[t]
\begin{center}
\includegraphics[width=8cm]{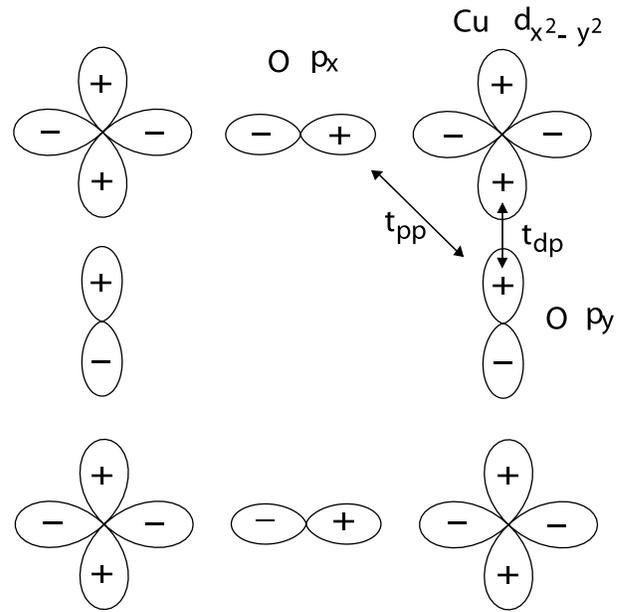}
\end{center}
\caption{
Lattice structure of d-p model. 
$t_{\rm dp}$ and $t_{\rm pp}$ are the hopping integrals. 
}
\label{fig:dpmodel}
\end{figure}

\begin{figure}[t]
\begin{center}
\includegraphics[width=4cm]{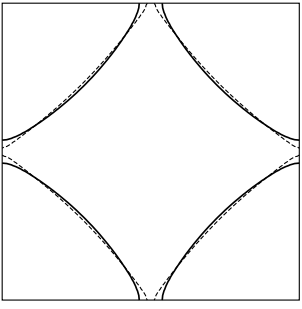}
\end{center}
\caption{
The unbroken and broken lines are the Fermi surfaces 
in the case of
$T=0.01$, $t_{\rm pp}=0.30$ and $n=4.90$ for 
$\varepsilon_d-\varepsilon_p=1.0$ and $\varepsilon_d-\varepsilon_p=3.0$, 
respectively. 
}
\label{fig:fs}
\end{figure}

Since the CuO$_2$ plane is essential for superconductivity in 
high-$\Tc$ cuprates, we can consider the lattice structure with only the 
CuO$_2$ network shown in Fig. \ref{fig:dpmodel}. 
This structure has the $d_{x^2-y^2}$ orbitals on 
Cu sites and the $p_x$ and $p_y$ orbitals on O sites in the primitive cell. 
For simplicity, we consider only the hopping integrals $t_{dp}$ and $t_{pp}$ 
shown in Fig. 
\ref{fig:dpmodel}. 
This is the so-called d-p model. 
In this study, we investigate in detail the nature of superconductivity 
in the d-p model. 
The model Hamiltonian is written as 
\begin{equation}
\begin{split}
H=H_0+H_{\rm int},
\end{split}
\end{equation}
where, 
\[
H_0=
\left(
\begin{array}{ccc}
d_{k\sigma}^{\dagger} & p_{k\sigma}^{x\dagger} & p_{k\sigma}^{y\dagger}
\end{array}
\right)
\left(
\begin{array}{ccc}
\varepsilon_d & \xi_k^x & \xi_k^y \\
\xi_k^x & \varepsilon_p & \xi^p_k \\
\xi_k^y & \xi^p_k & \varepsilon_p
\end{array}
\right)
\left(
\begin{array}{c}
d_{k\sigma}\\
p_{k\sigma}^{x}\\
p_{k\sigma}^{y}
\end{array}
\right).
\]
Here, $\varepsilon_d$ and $\varepsilon_p$ include the chemical potential 
$\mu$. The essential parameter is the level splitting 
$\varepsilon_d-\varepsilon_p$. 
The off-diagonal parts are given by 
\begin{equation}
\begin{split}
&\xi^{i}_k=-2t_{dp}\sin\frac{k_i}{2}, (i=x,y),\\
&\xi^{p}_k=4t_{pp}\sin\frac{k_x}{2}\sin\frac{k_y}{2}.
\end{split}
\end{equation}
The second term denotes the on-site Coulomb repulsion between the 
d-electrons, 
\begin{equation}
\begin{split}
H_{\rm int}=\frac{U}{N}\sum_k\sum_{q\neq 0}d^{\dagger}_{k+q\uparrow}d^{\dagger}_{k'-q\downarrow}d_{k'\downarrow}d_{k\uparrow}.
\end{split}
\end{equation}
We set $t_{dp}=1.0$ as an energy unit, and we also fix $t_{pp}=0.30$ and 
$n=4.90$ to reproduce the Fermi surfaces. 
In Fig. \ref{fig:fs}, we show the Fermi surfaces for
$\varepsilon_d-\varepsilon_p=1.0$ 
and $\varepsilon_d-\varepsilon_p=3.0$ in the 
case of 
$T=0.01$. 
We see that the Fermi surfaces coincide with those of the high-$\Tc$ cuprates, 
YBCO and LSCO, respectively. 
In this model, the electron number $n< 5.0$ represents the hole-doped case, 
and $n> 5.0$ represents the electron-doped case. 
We fix $T=0.01$ except when we discuss $\Tc$. 
\begin{figure}[t]
\begin{center}
\includegraphics[width=8cm]{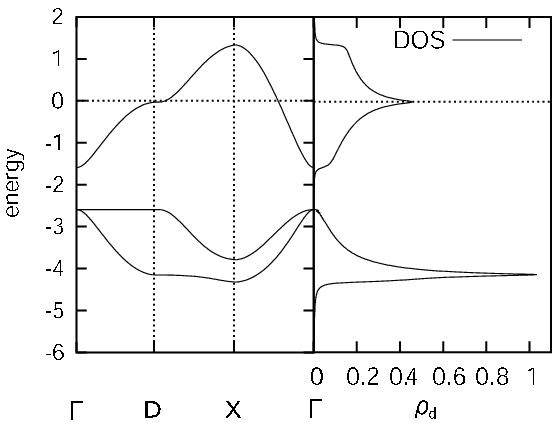}
\end{center}
\caption{
Band structure and density of states for d-electron $\rho_d(\varepsilon)$ 
in case of 
$\varepsilon_d-\varepsilon_p=1.0$. 
}
\label{fig:band1}
\end{figure}
\begin{figure}[t]
\begin{center}
\includegraphics[width=8cm]{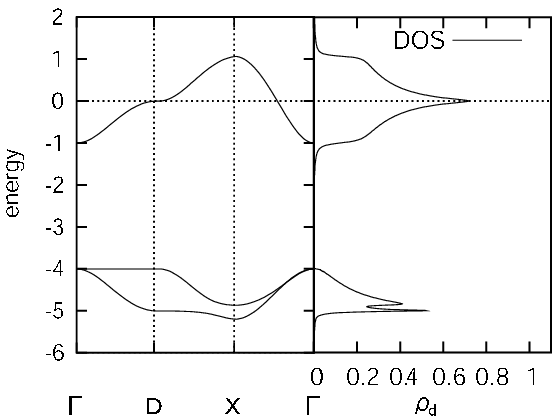}
\end{center}
\caption{
Band structure and density of states for d-electron $\rho_d(\varepsilon)$ 
in case of 
$\varepsilon_d-\varepsilon_p=3.0$. 
}
\label{fig:band3}
\end{figure}
Figs. \ref{fig:band1} and \ref{fig:band3} show the band structures and 
the density of states for d-electrons $\rho_d(\varepsilon)$ for 
$\varepsilon_d-\varepsilon_p=1.0$ and $\varepsilon_d-\varepsilon_p=3.0$, 
respectively. 
We see that when $\varepsilon_d-\varepsilon_p$ is large, 
$\rho_d(\varepsilon=0)$ at the Fermi level is large. \\
The bare Green's function $\hat{G}^{(0)}(k)$ is written as
\begin{equation}
\begin{split}
\hat{G}^{(0)}(k)=(i\omega_n\hat{1}-\hat{H}_0)^{-1}.
\end{split}
\end{equation}
We define the matrix elements of $\hat{G}^{(0)}(k)$ as 
\[
\hat{G}^{(0)}(k)=
\left(
\begin{array}{ccc}
G^{(0)}_{\rm dd}(k) & G^{(0)}_{\rm dp_x}(k) & G^{(0)}_{\rm dp_y}(k) \\
G^{(0)}_{\rm p_xd}(k) & G^{(0)}_{\rm p_xp_x}(k) & G^{(0)}_{\rm p_xp_y}(k) \\
G^{(0)}_{\rm p_yd}(k) & G^{(0)}_{\rm p_yp_x}(k) & G^{(0)}_{\rm p_yp_y}(k) 
\end{array}
\right).
\]
We need $G^{(0)}_{\rm dd}(k)$ to only describe the normal self-energy 
$\hat{\Sigma}(k)$ and 
the anomalous self-energy $\Delta(k)$, 
since the interaction exists only between 
the d-electrons.  
\[
\hat{\Sigma}(k)=
\left(
\begin{array}{ccc}
\Sigma_{\rm dd}(k) & 0 & 0 \\
0 & 0 & 0 \\
0 & 0 & 0 
\end{array}
\right).
\] 
The Green's function is given by 
\begin{equation}
\begin{split}
\hat{G}(k)=(i\omega_n\hat{1}-\hat{H}_0-\hat{\Sigma}(k))^{-1}.
\end{split}
\end{equation}
In this study, the chemical potential $\mu$ is determined 
so as to fix the total electron number, $n$, 
\begin{equation}
\begin{split}
n=2\frac{T}{N}\sum_k {\rm Tr}\,\hat{G}_0(k)=2\frac{T}{N}\sum_k {\rm Tr}\, \hat{G}(k).
\end{split}
\end{equation}

We apply the third-order perturbation theory
with respect to $U$. 
The normal self-energy is given by
\begin{equation}
\begin{split}
\Sigma_{\rm dd}(k)&= \frac{T}{N}\sum_{k'} [U^2 \chi_0(k-k') G_{\rm dd}^{(0)}(k') \\
& +U^3 \left( \chi_0^2(k-k')+\phi_0^2(k+k') \right) G_{\rm dd}^{(0)}(k')]. 
\end{split}
\end{equation}
Since the first-order term is constant, 
we can include the first-order term in $\varepsilon_d-\varepsilon_p$. 
Here, 
\begin{equation}
\begin{split}
&\chi_0(q)=-\frac{T}{N}\sum_{k} G_{\rm dd}^{(0)}(k)G_{\rm dd}^{(0)}(q+k), \\
&\phi_0(q)=-\frac{T}{N}\sum_{k}G_{\rm dd}^{(0)}(k)G_{\rm dd}^{(0)}(q-k).
\end{split}
\end{equation}
We also expand the effective pairing interaction up to the third-order terms 
with respect to $U$. \\
For the spin-singlet state, the effective pairing interaction is given by
\begin{equation}
\begin{split}
V(k;k')=V_{\rm RPA}(k;k')
+V_{\rm Vertex}(k;k'),
\end{split}
\end{equation}
where
\begin{equation}
\begin{split}
V_{\rm RPA}(k;k')=U+U^2\chi_0(k-k')
+2U^3\chi_0^2(k-k'), 
\end{split}
\end{equation}
and
\begin{equation}
\begin{split}
&V_{\rm Vertex}(k;k')=2(T/N)\Re\Big [\sum_{k_1}G_{\rm dd}^{(0)}(k_1) \\
&\times(\chi_0(k+k_1)-\phi_0(k+k_1))G_{\rm dd}^{(0)}(k+k_1-k')U^3\Big ].
\end{split}
\end{equation}
Here, $V_{\rm RPA}(k,k')$ is called the RPA terms and 
$V_{\rm Vertex}(k,k')$ is called the vertex 
corrections. 
Near the transition point, the anomalous self-energy $\Delta(k)$ satisfies 
the linearized \'Eliashberg equation,
\begin{equation}
\begin{split}
\lambda_{\rm max}\Delta(k)=-\frac{T}{N}\sum_{k'}V(k;k')|G_{\rm dd}(k')|^2\Delta(k'),
\end{split}
\end{equation}
where $\lambda_{\rm max}$ is the largest positive eigenvalue. 
In this equation, the temperature with $\lambda_{\rm max}=1$ 
corresponds to $T_{\rm c}$.
The symmetry of the superconductivity obtained here is the spin-singlet 
$d_{x^2-y^2}$ wave. 

\begin{figure}[t]
\begin{center}
\includegraphics[width=8cm]{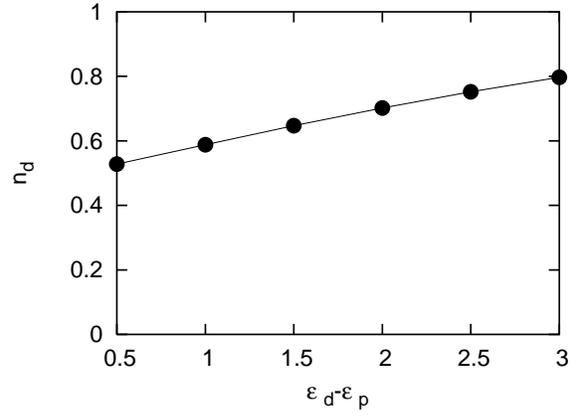}
\end{center}
\caption{
$\varepsilon_d-\varepsilon_p$ dependence of $n_d$. 
}
\label{fig:dpnd1}
\end{figure}

We take 64 $\times$ 64 $\ve{k}$-meshes for the 
first Brillouin zone and 2048 Matsubara frequencies in the numerical 
calculation. 
\begin{figure}[t]
\begin{center}
\includegraphics[width=8cm]{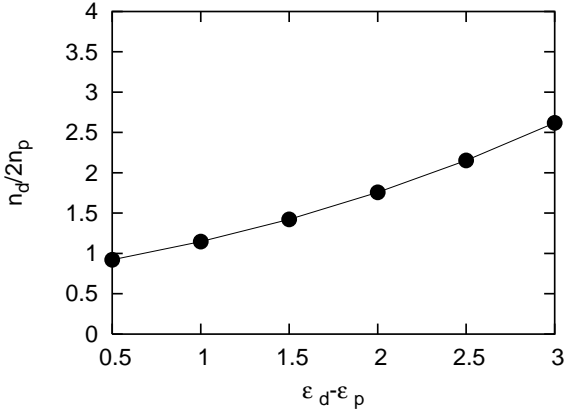}
\end{center}
\caption{
$\varepsilon_d-\varepsilon_p$ dependence of $n_d/2n_p$. 
}
\label{fig:dpratio1}
\end{figure}
First, we show the physical quantities in the unperturbed state. 
In Fig. \ref{fig:dpratio1}, we show the $\varepsilon_d-\varepsilon_p$ 
dependence of $n_d/2n_p$. 
When $\varepsilon_d-\varepsilon_p$ is large, $n_d/2n_p$ is large. 
Therefore, in the d-p model, the difference between LSCO 
and YBCO is represented as the difference of 
$\varepsilon_d-\varepsilon_p$. That is to say, 
$\varepsilon_d-\varepsilon_p$ is large for LSCO 
and small for YBCO, in this model. 
\begin{figure}[t]
\begin{center}
\includegraphics[width=8cm]{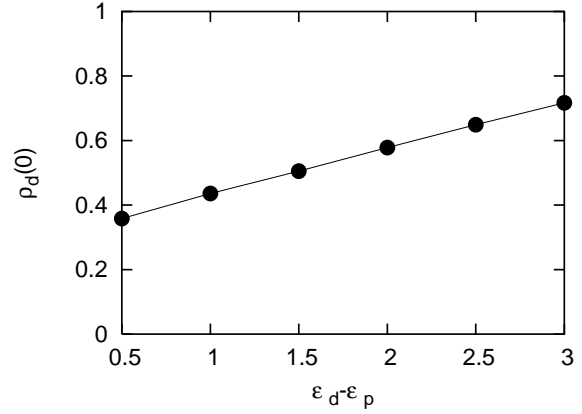}
\end{center}
\caption{
$\varepsilon_d-\varepsilon_p$ dependence of $\rho_d(0)$. 
}
\label{fig:dpdos1}
\end{figure}
In Figs. \ref{fig:dpdos1} and \ref{fig:dpnd1}, we show the 
$\varepsilon_d-\varepsilon_p$ dependences of $\rho_d(0)$ and $n_d$. 
When $\varepsilon_d-\varepsilon_p$ is large, these quantites also become 
large, and $n_d$ approaches unity. 
This indicates that LSCO is located in the relatively strongly correlated 
regime as compared with YBCO. 
\begin{figure}[t]
\begin{center}
\includegraphics[width=8cm]{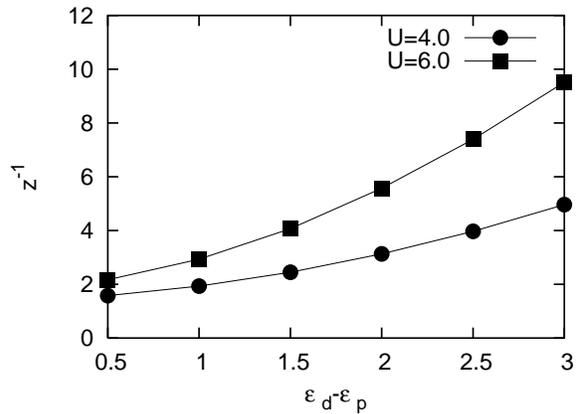}
\end{center}
\caption{
$\varepsilon_d-\varepsilon_p$ dependence of 
$z^{-1}=\langle z^{-1}(\ve{k})\rangle$. 
}
\label{fig:dpmass1}
\end{figure}

Now, let us evaluate the mass enhancement factor using the 
third-order perturbation theory. 
\begin{equation}
\begin{split}
z^{-1}(\ve{k})=\left (1-\frac{\partial \Sigma(\ve{k},\omega)}{\partial \omega}\right )_{\omega\rightarrow 0}. 
\end{split}
\end{equation}
Fig. \ref{fig:dpmass1} shows the $\varepsilon_d-\varepsilon_p$ 
dependence of the average $z^{-1}=\langle z^{-1}(\ve{k})\rangle$. 
Here, $\langle\cdots\rangle$ represents the average over the momentum space. 
With increasing $\varepsilon_d-\varepsilon_p$, the mass enhancement factor 
increases. 
This is consistent with the finding that the mass enhancement of LSCO seems to be 
larger than that of YBCO, as mentioned above. ~\cite{rf:Loram,rf:Momono} 
This is due to the difference in $\rho_d(0)$. 
From another viewpoint, 
when the value of $\varepsilon_d-\varepsilon_p$ is large, $n_d$ 
approaches unity, and the electron mass is strongly enhanced by the strong 
electron correlation, similarly to the Mott transition. 
Such behavior has been markedly observed in the Ce-based heavy 
fermion systems. 

\begin{figure}[t]
\begin{center}
\includegraphics[width=8cm]{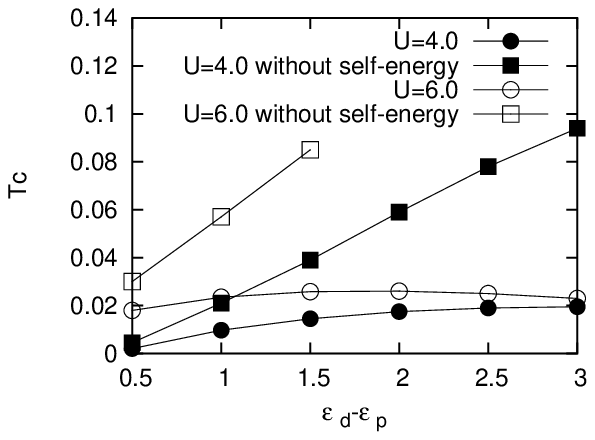}
\end{center}
\caption{
$\varepsilon_d-\varepsilon_p$ dependence of $\Tc$. 
}
\label{fig:dptc1}
\end{figure}

Next, we discuss the superconducting transition temperature $\Tc$. 
From the experimental results, for large $\varepsilon_d-\varepsilon_p$, 
$\Tc$ is relatively low. 
In Fig. \ref{fig:dptc1}, we show the $\varepsilon_d-\varepsilon_p$ 
dependence of $\Tc$. 
$\Tc$ for $U=4.0$ increases as a function of $\varepsilon_d-\varepsilon_p$, 
and for $U=6.0$ is almost unchanged. Although the normal 
self-energy correction 
effect markedly suppresses $\Tc$, 
we cannot derive the results observed in the experiments within the present 
calculation. 
On the basis of the following reason, 
however, we can expect that by calculating the higher-order terms, 
we will obtain a better dependence of $\Tc$ on $\varepsilon_d-\varepsilon_p$. 
In Fig. \ref{fig:dptc1}, the filled (open) 
circles and squares denote $\Tc$ for $U=4.0$ ($U=6.0$) with and 
without the normal self-energy correction, respectively. 
The difference is mainly due to the effect of the mass renormalization. 
The larger $\varepsilon_d-\varepsilon_p$ is, the more this effect increases. 
This tendency is considered to be more marked in the higher-order terms of 
the normal self-energy. 
On the other hand, higher-order terms of the pairing interaction were evaluated by 
Nomura {\it et al.} in the Hubbard model. 
The results show that the convergency for 
the pairing interaction is very good for the spin-singlet pairing 
near the half-filling.~\cite{rf:Nomura} 
Namely, the contribution of the fourth-order terms to the pairing interaction 
is small compared with those of second-order and third-order terms. 
Here, we can expect the same trend. 
Thus, with the inclusion of the fourth-order terms 
in the \'Eliashberg equation, 
$\Tc$ will become relatively low for large $\varepsilon_d-\varepsilon_p$. 
This prediction is an important problem to be confirmed in the future. 

In conclusion, 
we have investigated the mass enhancement factor and the superconducting 
transition temperature in the d-p model for the high-$\Tc$ cuprates. 
We have solved the \'Eliashberg equation using the third-order perturbation theory 
with respect to the on-site repulsion $U$. 
We find that when $\varepsilon_d-\varepsilon_p$ is large, 
the mass enhancement factor becomes large and $\Tc$ tends to be suppressed 
owing to the difference in $\rho_d(0)$. 
From another viewpoint, when the d-hole number approaches 
unity, the electron correlation 
between d-holes (electrons) becomes strong and the effective mass increases. 
In fact, LSCO with d-hole number near unity shows 
strong mass enhancement . 
Here, we consider that $\Tc$ is given by the renormalization factor $z$ and
$\Tc^{\ast}$ as  
\begin{equation}
\begin{split}
\Tc\simeq z\Tc^{\ast},
\end{split}
\end{equation}
where $\Tc^{\ast}$ is the critical temperature determined by the calculation 
without any renormalization due to the normal self-energy correction. 
LSCO possesses small $z$ and exhibits low $\Tc$. Also, in the 
heavy fermion systems, when the number of 
f-electrons approaches unity, the effective mass is large and $\Tc$ is 
suppressed. 
This is the important unified
theory which holds for all the strongly correlated electron systems from 
cuprates to heavy fermions.  
Thus, our calculation shows that, in order to systematically discuss 
the physical quantities such as 
$\Tc$ and the 
mass enhancement factor, we need to use the d-p model. The Hubbard 
Hamiltonian is insufficient to represent the difference among 
material systems. \\

{\bf Acknowledgments}\\

Numerical calculation in this work was carried out at 
the Yukawa Institute Computer Facility.


\end{document}